\documentclass[aps,prd,nofootinbib,showkeys,twocolumn,floatfix]{revtex4}

\usepackage{amssymb}
\usepackage{amsmath}
\usepackage{amsfonts}
\usepackage{graphicx}
\usepackage{color}
\usepackage{xspace}
\usepackage{ulem}
\usepackage{lscape}
\usepackage{ wasysym }

\usepackage{rotating}

\def\gsim{\raise0.3ex\hbox{$\;>$\kern-0.75em\raise-1.1ex\hbox{$\sim\;$}}}
\def\lsim{\raise0.3ex\hbox{$\;<$\kern-0.75em\raise-1.1ex\hbox{$\sim\;$}}}

\def\znbb{0\nu\beta\beta}

\newcommand{\ba}[1]{\begin{eqnarray} \label{(#1)}}
\newcommand{\ea}{\end{eqnarray}}

\newcommand{\AddrAHEP}{
  {\it AHEP Group, Instituto de F\'{\i}sica Corpuscular --
    C.S.I.C./Universitat de Val{\`e}ncia \\
    Edificio de Institutos de Paterna, Apartado 22085,
  E--46071 Val{\`e}ncia, Spain}}

\newcommand{\AddrUFSM}{
Universidad T\'ecnica Federico Santa Mar\'\i a, \\ 
Centro-Cient\'\i fico-Tecnol\'{o}gico de Valpara\'\i so, \\ 
Casilla 110-V, Valpara\'\i so,  Chile}

\def\gsim{\raise0.3ex\hbox{$\;>$\kern-0.75em\raise-1.1ex\hbox{$\sim\;$}}}
\def\lsim{\raise0.3ex\hbox{$\;<$\kern-0.75em\raise-1.1ex\hbox{$\sim\;$}}}

\newcommand{\nn}{\nonumber}

\begin{document}

\preprint{IFIC/13-94}  

\title{Heavy neutrino searches at the LHC with displaced vertices}

\author{J.C. Helo} \email{juan.heloherrera@gmail.com}\affiliation{\AddrUFSM}
\author{M. Hirsch} \email{mahirsch@ific.uv.es}\affiliation{\AddrAHEP}
\author{S.G. Kovalenko}\email{Sergey.Kovalenko@usm.cl}\affiliation{\AddrUFSM}

\keywords{neutrino masses and mixing; LHC}

\pacs{14.60.Pq, 12.60.Jv, 14.80.Cp}

\begin{abstract}
Sterile neutrinos with masses in the range (1-100) GeV, have been
searched for in a variety of experiments. Here, we discuss the
prospects to search for sterile neutrinos at the LHC using displaced
vertices. Two different cases are discussed: (i) the standard model
extended with sterile neutrinos and (ii) right-handed neutrinos in a
left-right symmetric extension of the standard model. A dedicated
displaced vertex search will allow to probe parts of the parameter
space not accessible to other searches, but will require large
luminosity in both cases.

\end{abstract}

\maketitle

\section{Introduction}

Searches for heavy neutral leptons have been carried out in a variety
of experiments \cite{Beringer:1900zz}.  Lacking any positive evidence
so far, upper limits on the mixing of these ``sterile neutrinos'' with
the ordinary, light, active neutrinos have been derived as function of
the neutrino mass, with the most stringent limits for the mass range
$m_N = [3,80]$ GeV on $|V_{\mu 4}|^2 \lsim 2 \times 10^{-5}$ coming
still from the DELPHI experiment at LEP \cite{Abreu:1996pa}, while
$|V_{e 4}|^2 \lsim 10^{-7} \times \frac{m_N}{\rm GeV}$ has been
calculated from the absence of neutrinoless double beta decay
\cite{Benes:2005hn}.  In the mass range $m_N = [0.1,3]$ GeV 
meson decays give the most stringent limits, due to a resonant 
enhancement of the decay rate, if the sterile neutrino mass
$m_N$ is close to its mass-shell \cite{Helo:2010cw}. For a 
complete list of limits on heavy sterile (Majorana) neutrino 
searches see \cite{Atre:2009rg}.

In this paper, we discuss possible searches for sterile Majorana
neutrinos with masses $m_N = [3,80]$ GeV at the LHC.  The classical
signal for lepton number violation at a hadron collider is like-sign
di-leptons with (two) jets and no missing energy \cite{Keung:1983uu},
although also tri-lepton final states as possible signal have been
suggested \cite{delAguila:2008cj}. We instead discuss displaced
vertices as a possibility to search for sterile neutrinos at the LHC.
The advantage of a displaced vertex search is that for decay lengths
of the order of very roughly $L = [10^{-3},1]$ m there is little (or
no) standard model background. Thus, in principle, a discovery could
be claimed with as few as (3-5) events.

Below, we will consider two different setups in which heavy neutral
leptons could occur. (i) Motivated by the seesaw mechanism, the
standard model extended by the addition of a sterile Majorana
neutrino.  And, (ii) the (minimal) left-right symmetric extension of
the standard model.

Neutrino oscillation experiments have shown that neutrinos have
non-zero masses. For the current status of oscillation data see, for
example \cite{Tortola:2012te}. In the classical seesaw mechanism
\cite{Minkowski:1977sc,seesaw,Mohapatra:1979ia,Schechter:1980gr} the
smallness of the neutrino mass is due to the large mass of the
right-handed neutrinos, expressed in the famous seesaw relation as
$m_{\nu} \propto \frac{m_D^2}{M_M}$. For Yukawa couplings of order $Y
\sim (10^{-2}-1)$ current neutrino data then put $M_M$ in the range
$M_M \simeq (10^{11}-10^{15})$ GeV, forever in the realm of
theoretical speculation. However, for smaller Yukawa couplings, say of
the order $Y \sim (10^{-7}-10^{-6})$, sterile neutrinos with masses
$m_N = (10-1000)$ GeV could exist. Note that the naive expectation for
heavy-light neutrino mixing in the seesaw is $V \propto \sqrt{m_{\nu}/M_N}$, 
i.e. $|V_{l 4}|^2 \simeq 5 \times 10^{-14}
(\frac{m_{\nu}}{\rm 0.05 eV})(\frac{\rm 1 TeV}{M_N})$. Obviously, 
none of the limits on heavy lepton mixing discussed above comes 
even close to this expectation.

Expectations for a discovery are much higher for left-right (L-R)
symmetric extensions of the standard model
\cite{Pati:1974yy,Mohapatra:1974gc}. If the right-handed neutrinos,
necessarily present in L-R models, have a Majorana mass
\cite{Mohapatra:1980yp}, they appear in accelerators as ``sterile''
heavy objects and their Majorana nature can be probed by searching for
like-sign di-leptons \cite{Keung:1983uu}. For $m_{W_R} >> m_{W_L}$,
but $m_{N} << m_{W_R}$ the sterile neutrino N can be produced via an
effective operator $pp \to e^{\pm} + N$, with $N$ decaying to $N\to
e^{\pm}jj$. Different from the case of the standard model plus sterile
neutrino, here the production cross section and the decay length are
not suppressed by the small mixing $|V_{l 4}|^2$, but 
depend mostly on the unknown mass of the heavy $W_R$.

The rest of this paper is organized as follows. In the next section 
we briefly summarize the main formulas for production and decay 
of sterile neutrinos for our two models. Section \ref{sect:dsplcd} 
then discusses numerical results for displaced vertex searches, 
first in section \ref{sect:sterile} for SM + sterile neutrino 
and in section \ref{sect:LR} for the case of LR symmetry. 
We then close the paper with a short discussion.

\section{Theoretical setup}
\label{sect:theo}

Here we consider two setups with sterile neutrinos: (i) an extension
of the SM field contents with a non-fixed number $i$ of the SM singlet
fermions $\nu_{Ri}$ of unspecified origin and (ii) the Left-Right
(L-R) symmetric model having three right-handed neutrinos $\nu_{Ri}$
the L-R partners of the usual left-handed neutrinos
$\nu_{e,\mu,\tau}$.

\subsection{ A generic sterile neutrino scenario.}
Adding $n$ species of SM singlet right-handed neutrinos
$\nu^{\,\prime}_{R\alpha}=(\nu^{\,\prime}_{R1},...\nu^{\,\prime}_{Rn})$
allows together with the three left-handed weak doublet neutrinos
$\nu^{\,\prime}_{L\beta} =
(\nu^{\,\prime}_{Le},\nu^{\,\prime}_{L\mu},\nu^{\,\prime}_{L\tau})$
one to construct a mass term
\begin{eqnarray}\label{Mass-Term}
&&-\frac{1}{2} \overline{\nu^{\,\prime}} {\cal M}^{(\nu)} \nu^{\,\prime
c} + \mbox{h.c.} =\\
\nonumber
&&-\frac{1}{2} (\overline{\nu^{\,\prime}}_{_L},
\overline{\nu_{_R}^{\,\prime c}}) \left(\begin{array}{cc}
{\cal M}_L & {\cal M}_D \\
{\cal M}^T_D  & {\cal M}_R \end{array}\right) \left(\begin{array}{c}
\nu_{_L}^{\,\prime c} \\
\nonumber
\nu^{\,\prime}_{_R}\end{array}\right) + \mbox{h.c.} \\
\nonumber
&=&-\frac{1}{2} (\sum_{i=1}^{3} m_{\nu_i} \overline{\nu^c_i}\nu_i
+\sum_{j=1}^{n} m_{\nu_j} \overline{\nu^c_j}\nu_j  )+ \mbox{h.c.}
\end{eqnarray} 
where  ${\cal M}_L, {\cal M}_R$ are $3\times 3$ and $n\times n$
symmetric Majorana mass matrices, and ${M}_D$ is a $3\times n$ Dirac
type matrix.  In the seesaw language the entry ${\cal M}_L \neq 0$
is due to the seesaw type-II mechanism. Below we study a generic scenario without 
specifying concrete mechanism generating the mass matrix in Eq. (\ref{Mass-Term}).

A unitary rotation
 \begin{eqnarray}\label{rot-1}
 && V^T {\cal M}^{(\nu)}V =
\textrm{Diag}\{m_{\nu 1}, \cdots ,m_{\nu_{3+n}}\}
\label{mass-eigen}
\end{eqnarray}
leads to $3+n$ Majorana neutrinos with masses $m_{\nu_1}, \cdots,
m_{\nu_{3+n}}$.  The matrix $V_{\alpha k}$ is the neutrino mixing
matrix.  This spectrum must contain the three very light neutrinos
with different masses and dominated by the active flavors
$\nu_{\alpha}$ ($\alpha = e, \mu, \tau$) and $n$ heavy states $N_{Ri}$
with the masses $M_{NR_{i}}$ and the active neutrino admixtures
$V_{\alpha N_{Ri}}$ subject to the
experimental constraints discussed in the introduction (see, for
instance \cite{Beringer:1900zz,Atre:2009rg}).
In what follows we consider the case of only one heavy sterile
neutrino $N$.  Thus there are four parameters characterizing 
the scenario in question: the neutrino mass $m_{N}$ and its mixing 
$V_{\alpha N}$ with the active flavors $\alpha=e, \mu, \tau$.

The heavy sterile neutrino Charged (CC) and Neutral
Current (NC) interactions are
\begin{eqnarray}\label{CC-NC}
{\cal L} &=& \frac{g_{L}}{\sqrt{2}}\, 
\left( V^{L}_{l i}\ \bar l \gamma^{\mu} P_L \nu_{i} 
 +  V_{l N}\ \bar l \gamma^{\mu} P_L N\right) W^-_{L \mu}
+  \\
\nonumber
&+&\frac{g}{2 \cos\theta_W}\ \sum_{\alpha, i, j}V^{L}_{\alpha i} V_{\alpha N}^*  
\bar{N} \gamma^{\mu} P_L \nu_{i}\  
Z_{\mu},
\end{eqnarray}
where $i=1,2,3$. The left-handed sector neutrino mixing matrix
$V^{L}$, that appears in neutrino oscillations , coincides with the
PMNS matrix.  The matrix $V_{\alpha N}$ describes the mixing between
heavy and light neutrino sectors. The heavy neutrino $N$ total decay
rate $\Gamma_{N}$ in the mass range $m_{N}\subset$[1,80] GeV, used 
in the next section, is given in Appendix.

The half-life $T_{1/2}$ for neutrinoless double beta
($0\nu\beta\beta$) decay in this framework is
\begin{eqnarray}\label{NLDBD-SNU}
T^{-1}_{1/2} =  G_{01} |{\cal{M}}_{N}|^{2}
 \left|\frac{V^{2}_{eN}}{m_{N}}\right|^{2}
\end{eqnarray} 
where $G_{01}$ and ${\cal M}_{N}$ are the phase space factor and
nuclear matrix element. Their values can be found in, for instance, 
the recent review \cite{Deppisch:2012nb}.

\subsection{Minimal left-right symmetric extension of the standard model.} 
This is a significantly more restrictive framework \mbox{\cite{Pati:1974yy,Mohapatra:1974gc,Mohapatra:1980yp}} based on the
gauge group $SU(2)_{L}\times SU(2)_{R} \times U(1)_{(B-L)}$ with the
gauge couplings $g_{L}, g_{R}, g_{1}$ and the LR symmetric assignment
of the quarks and leptons: $Q_{L,R} = (u, d)_{L,R}$ and $L_{L,R} =
(\nu, l)_{L,R}$.  The gauge symmetry is broken to the SM symmetry at
some scale $M_{R}$, at least somewhat larger than the electroweak
symmetry breaking scale.  Now there are three heavy neutrino mass
eigenstates $N$ one per family.

The CC and NC interactions relevant for our analysis are
\begin{eqnarray}\label{CC-NC-LR}
{\cal L} &=& \frac{g_{R}}{\sqrt{2}}  
\left(\bar{d}  \gamma^{\mu} P_R u 
+  V^{R}_{l N}\cdot \bar l \gamma^{\mu} P_R N\  \right) W^-_{R \mu} 
+  \\
\nonumber
&+&\frac{g_{R}}{\sqrt{1-\tan^{2}\theta_{W} (g_{L}/g_{R})^{2}}}\times\\
\nonumber
&& Z_{LR}^{\mu}\bar{f} \gamma_{\mu}\left[ T_{3R} +  \tan^{2}\theta_{W} (g_{L}/g_{R})^{2}\left(T_{3L} - Q\right)\right] f  
\end{eqnarray}
where $V^{R}$ is the right-handed sector neutrino mixing matrix.  The
charged $W_{L,R}$-boson states are expressed in terms of the mass
eigenstates as:
\begin{eqnarray}\label{W-MES}
W^{-}_{L} &=& \cos\zeta\cdot  W^{-}_{1} - \sin\zeta\cdot W^{-}_{2},\\
\nonumber
W^{-}_{R} &=& \sin\zeta\cdot W^{-}_{1} + \cos\zeta\cdot W^{-}_{2},
\end{eqnarray}
where the $W_{L}-W_{R}$ mixing angle in a good approximation 
\cite{MohapatraPal}  is 
\begin{eqnarray}\label{WL-WR-mix}
\tan 2\zeta &=& \frac{2 g_{L}g_{R} M^{2}_{W_{L}} \cdot  
\sin 2\beta}{g_{R}^{2} M_{W_{L}}^{2} + g_{L}^{2}(M_{W_{R}}^{2} - M_{W_{L}}^{2})} \\
\nonumber
&\approx&  2\  \frac{g_{R}}{g_{L}} \frac{M^{2}_{W_{L}}}{M^{2}_{W_{R}}} \sin 2\beta .
\end{eqnarray} 
Here $\tan\beta = \kappa^{\prime}/\kappa$ is the ratio of the two
vev's of the bidoublet Higgs $\Phi$.  In the last step we used 
$M_{W_{R}}\gg M_{W_{L}}$.  For convenience we denoted the masses of
$W^{-}_{1,2}$ as $M_{W_{L}}$ and $M_{W_{R}}$ respectively.  
Maximal $W_{L}-W_{R}$ mixing is reached for $\kappa = \kappa^{\prime}$
when $\sin 2\beta = 1$.  Imposing the perturbativity condition
$g_{R}^{2}/4\pi \leq 1$ and using $g_{L} = g(M_{Z}) \approx 0.64935$
\cite{Beringer:1900zz} we find the upper bound
\begin{eqnarray}\label{WL-WR-mix-max}
\tan 2\zeta &\leq& 
\frac{4 \sqrt{\pi}}{g(M_{Z})} \frac{M_{W_{L}}^{2}}{M_{W_{R}}^{2}}.
\end{eqnarray}  
For the total decay of each of the three heavy neutrinos of the L-R
models in the mass range \mbox{$m_{N}\subset$[3,80] GeV} the neutral
current contribution is negligible and we approximately have:
\begin{eqnarray}\label{Dec-Rate-RR}
\Gamma_{N} 
\approx \frac{3  G^{2}_{F}}{32 \pi^{3}} m_{N}^{5} 
\left(\frac{M_{W_{L}}}{M_{W_{R}}}  \frac{g_{R}}{g_{L}}\right)^{4} 
\left[1 + \sin^{2} 2\beta \right] \sum_{l} |V^{R}_{l N}|^{2}, \ \ 
\end{eqnarray}
where we neglected the masses of all the final state particles.

The half-life $T_{1/2}$ for neutrinoless double beta ($0\nu\beta\beta$) 
decay via heavy $W_R$ and heavy $N$ exchange is given by:
\begin{eqnarray}\label{NLDBD-LR}
T^{-1}_{1/2} =  G_{01} |{\cal{M}}_{N}|^{2} 
\left| \left(V^{R}_{eN}\right)^{2}  \frac{m_{p}}{m_{N}}  
\frac{M^{4}_{W_{L}}}{M^{4}_{W_{R}}} \frac{g^{4}_{R}}{g^{4}_{L}}
\right|^{2}
\end{eqnarray} 
In the numerical section we will consider a simplified case with 
only one heavy neutrino in the relevant mass range.

\section{Displaced vertex searches}
\label{sect:dsplcd}

We have implemented the two models discussed above in CALCHEP
\cite{Pukhov:2004ca} for the calculation of the cross section.  We
have compared our calculation for the sterile neutrino with cross
sections available in the literature
\cite{Atre:2009rg,delAguila:2008cj} and found good agreement.  
  We will first discuss the theoretically simpler
case of SM plus sterile neutrino.

\subsection{Sterile neutrino case}
\label{sect:sterile}

In the SM plus sterile neutrino scenario, the decay length of 
the heavy neutrino can be written as a function of two unknown 
parameters:
\begin{equation}\label{eq:decst}
L = c \bar \gamma \tau_{_{N}} \simeq  3.7  \bar \gamma   \left(\frac{\rm GeV}{m_N}\right)^5
\left(\frac{10^{-4}}{|V_{l4}|^2}\right)  [m]
\end{equation}
Here, $l=e,\mu$ and $\bar \gamma =\bar E_N /m_N$, where $\bar E_N$ is
the average energy of the heavy neutrino $N$ which we calculated
numerically. The heavy neutrino half-life is denoted by $\tau_{_{N}}$.
Decay lengths testable at the LHC are
roughly of the order of $L \simeq (10^{-3}-1)$ m and thus, sterile
neutrino masses between $m_N \simeq [1,30]$ GeV could lead to
measurable displaced vertices.

The cross section for sterile neutrino production at 14 TeV 
can be separated into a bare cross sections times a factor 
containing the neutrino mixing \cite{Atre:2009rg}. The bare 
cross section is huge, reaching several nb for neutrino masses 
below $m_W$. This allows, in principle, to constrain heavy neutrino 
mixing to impressively small values. However, the resulting 
events from production and decay of sterile neutrino produced 
by an off-shell W-boson are relatively soft. Correspondingly 
we have found that our results do very sensitively depend on 
the cuts used in the analysis. 

\begin{figure}[htbp]
\begin{minipage}[b]{.95\linewidth}
\includegraphics[width=\linewidth]{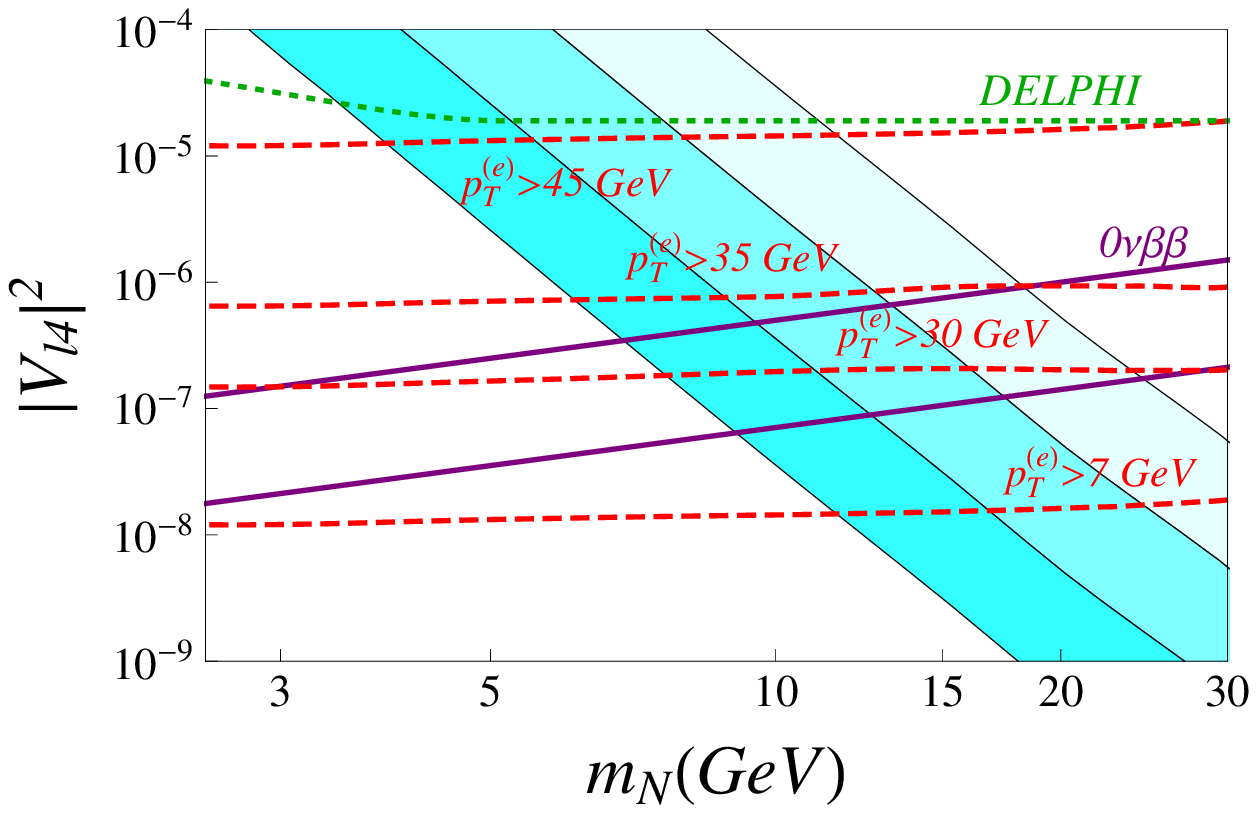}
\end{minipage}
\\
\begin{minipage}[b]{.95\linewidth}
\includegraphics[width=\linewidth]{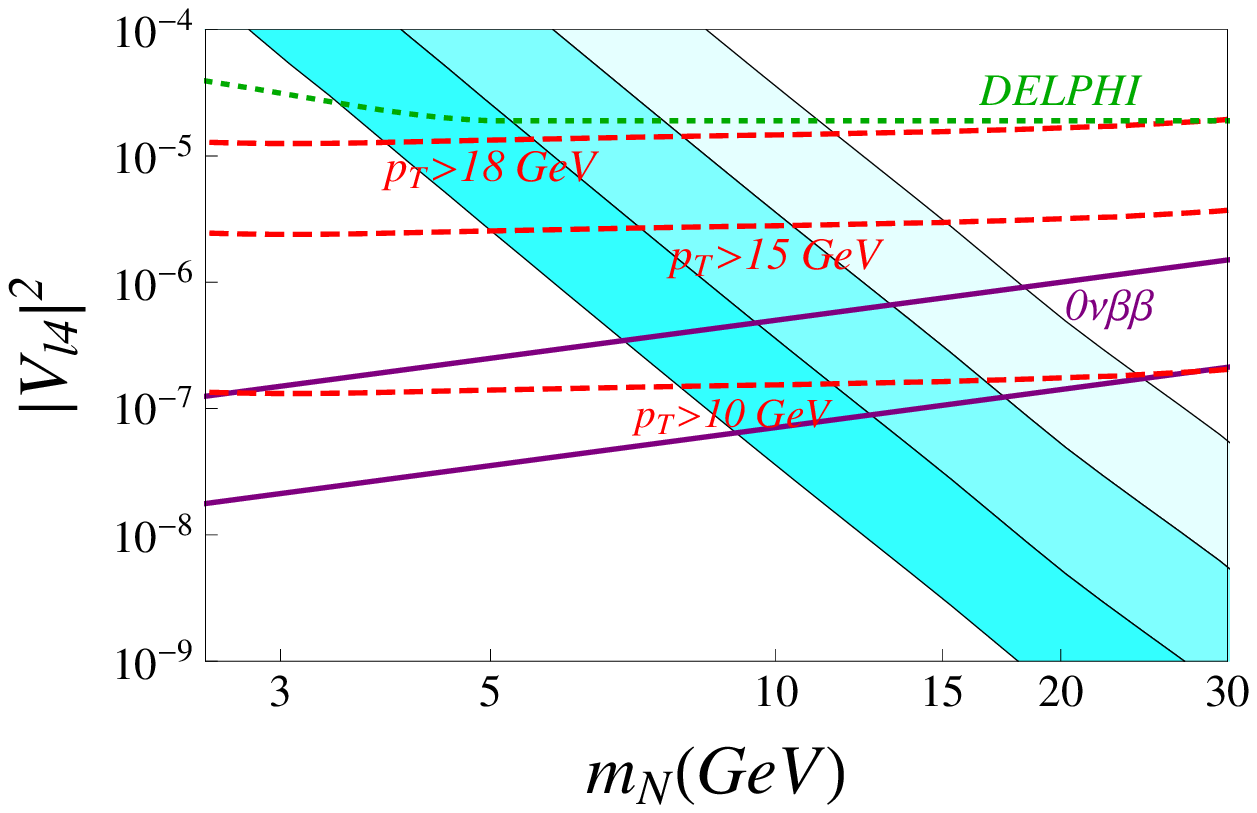}
\end{minipage}
\vspace{0.2 cm}
\caption{Constraints on sterile neutrinos from DELPHI
  \cite{Abreu:1996pa}, compared with double beta decay and the region
  in parameter space where a displaced vertex search at LHC will be
  sensitive.
   (a) top, cuts: $p_T^{e_1} > 30$ GeV, $p_T^{e_2} > 7$ GeV,
  $30$ GeV, $35$ GeV, $45$ GeV and $|\eta^{e}| < 2.5$. Luminosity:
  ${\cal L} = 300$ fb$^{-1}$. (b) bottom using $p_T^{e_2} > 7$ GeV,
  $p_T^{j} > 10$, $15$ and $20$ GeV, $p_T^{e_1} > 30 GeV$ and
  $|\eta^{e,j}| < 2.5$.   The limit from double beta decay applies only to
$l=e$, see text.  }
\label{fig:Sterile}  
\end{figure}

The lepton-based triggers used by the ATLAS collaboration in the 2012
run were \cite{VMitsou2013}: For two electrons (muons), either both
electrons (muons) have to have a $p_T\ge 12$ GeV ($p_T\ge 13$ GeV) or
the leading electron (muon) has to have $p_T\ge 24$ GeV ($p_T\ge 18$
GeV) with the second electron required to have $p_T\ge 7$ GeV ($p_T\ge
8$ GeV). We have then calculated possible constraints on $|V_{l4}|^2$
as function of the heavy neutrino mass $m_N$ for a variety of cuts,
motivated by these experimental cuts, but with variations as given
below to check the future reach that could be obtained under different
assumptions. The results are shown in fig. (\ref{fig:Sterile}).

Fig. (\ref{fig:Sterile}) shows constraints on $|V_{l4}|^2$ obtained by
the DELPHI collaboration \cite{Abreu:1996pa}, together with two limits
from non-observation of neutrinoless double beta decay
($0\nu\beta\beta$): The weaker limit shown is derived from the latest
search result by the GERDA collaboration ($T_{1/2} \ge 2.0 \times
10^{25}$ ys for $^{76}$Ge \cite{Agostini:2013mzu}), while the line at smaller
values of $|V_{l4}|^2$ is the expected sensitivity for $T_{1/2} \ge
10^{27}$ ys. Note that the limit from DELPHI applies to both
$l=e,\mu$, while the limit from double beta decay applies only to
$l=e$. The coloured bands show the region that can be probed at the
LHC in a 14 TeV run by a displaced vertex search. The three bands
correspond to decay lengths of order $L=1$ mm, $L=0.01$, $0.1$ and
$L=1$ m, with the smallest length for largest values of $|V_{l4}|^2$.
In this plot we assume a luminosity of ${\cal L} = 300$ fb$^{-1}$.

The red dashed lines are the expected sensitivity for the LHC 
assuming less than five signal events as the experimental upper limit.  
Different cuts on energies and $p_T$ have then be used to estimate the
sensitivty of the LHC. Consider the top panel first. Here, $|\eta^{e}|
< 2.5$ and the $p_T$ of the first electron is required to be
$p_T^{e_1} > 30$ GeV, while for the second electron (the one coming
from the displaced vertex, not necessarily the softer of the two
electrons) we require different values of $p_T> 7$, $30$, $35$ and
$45$ GeV.  It is clear that lowering the cut on the displaced vertex
electron as much as possible is absolutely essential in this
search. However, the plot shown in the top of fig. (\ref{fig:Sterile})
does not show a (completely) realistic situation, since no cut on the
jet energy was applied. Thus, while these events would show clearly
two electrons, with one coming from the displaced vertex, the hadronic
activity at the displaced vertex might be too soft to allow for jet
reconstruction.  For a more realistic estimate we thus show in the
same figure in the bottom panel the reach of the LHC, requiring
$p_T^{e_2} > 7$ GeV, $p_T^{j} > 10$, $15$ and $20$ GeV, $p_T^{e_1} >
30 GeV$ and $|\eta^{e,j}| < 2.5$. The additional cut on the jet $p_T$
again leads to a rapid loss of sensitivity, thus for this search to be
effective, experimentalists will have to lower the threshold for jet
search in displaced vertices as much as possible.

\begin{figure}[htbp]
\begin{minipage}[b]{.95\linewidth}
\includegraphics[width=\linewidth]{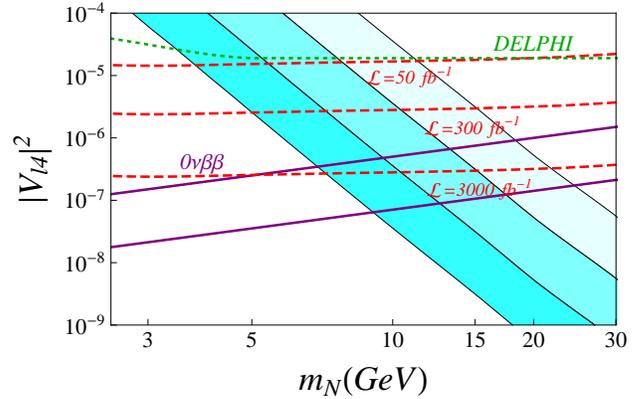}
\end{minipage}
\vspace{0.2 cm}
\caption{Constraints on sterile neutrinos from DELPHI
  \cite{Abreu:1996pa}, compared with double beta decay and the region
  in parameter space where displaced vertex search at LHC will be
  sensitive. Cuts: $p_T^{e_1} > 30$ GeV, $p_T^{e_2} > 7$ GeV, $p_T^{j}
  > 15 GeV$ and $|\eta| < 2.5$. Lines for different values for
  luminosity: ${\cal L} = 50$, $300$ and $3000$ fb$^{-1}$.  The limit from double beta decay applies only to
$l=e$, see text.}
\label{fig:Sterile2}  
\end{figure}

In fig. (\ref{fig:Sterile2}) we then show the sensitivity of the LHC
in the same plane as fig. (\ref{fig:Sterile}), but now for fixed
values of the cuts and for different assumed values of the luminosity:
${\cal L} = 50$, $300$ and $3000$ fb$^{-1}$. LHC could probe 
for $l=\mu$ so far unexplored ranges of $|V_{l4}|^2$ for luminosities 
as small as ${\cal L} = 50$ fb$^{-1}$. To do better than the current 
limit from $\znbb$ on $|V_{e4}|^2$, very large luminosities or 
significantly lower $p_T$ cuts will be necessary.

\subsection{Left-right symmetric model}
\label{sect:LR}

Now we will discuss the results for the left-right symmetric model. 
For the sake of simplicity we will start our discussion assuming 
``manifest'' L-R symmetry, i.e. $g_R=g_L$. In the LRSM the decay
length can be written as function of the two masses $m_N$ and
$m_{W_R}$:
\begin{equation}\label{eq:decLR}
L = c \bar \gamma \tau_{_{N}} \simeq  0.12 \ \bar \gamma  \ \left(\frac{10 \rm  GeV}{m_N}\right)^5
\left(\frac{m_{W_R}}{1 \ \rm  TeV}\right)^4  [mm]
\end{equation}
Here $\bar \gamma = \bar E_N /m_N$ which we calculated numerically. 
Here and in the following plots, we have neglected
effects of $W_L-W_R$ mixing, since $\sin 2\beta \le 1$. For the
extreme case of $\sin 2\beta$ approaching one, decay lengths will be a
factor of two smaller than shown in the figures, compare
eq. (\ref{Dec-Rate-RR}). 

\begin{figure}[htbp]
\begin{minipage}[b]{.95\linewidth}
\includegraphics[width=\linewidth]{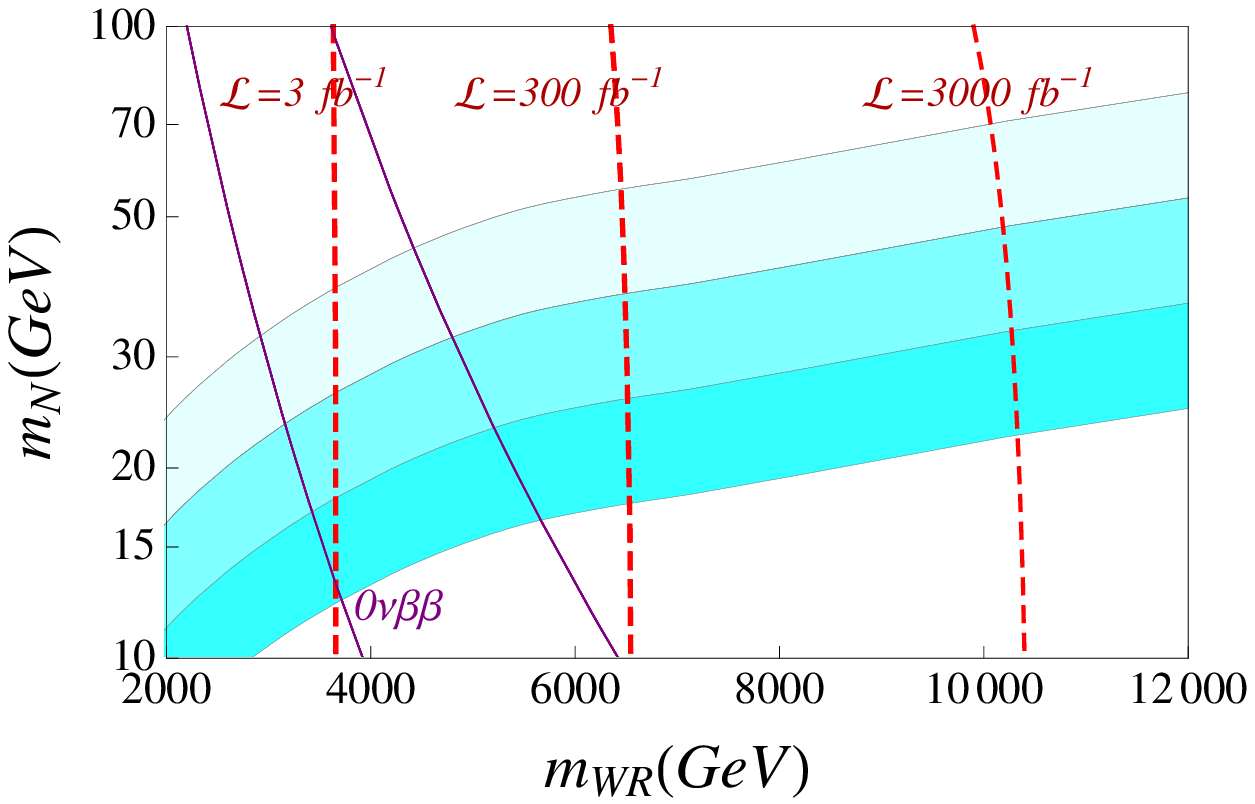}
\end{minipage}
\\
\begin{minipage}[b]{.95\linewidth}
\vspace{0pt}
\includegraphics[width=\linewidth]{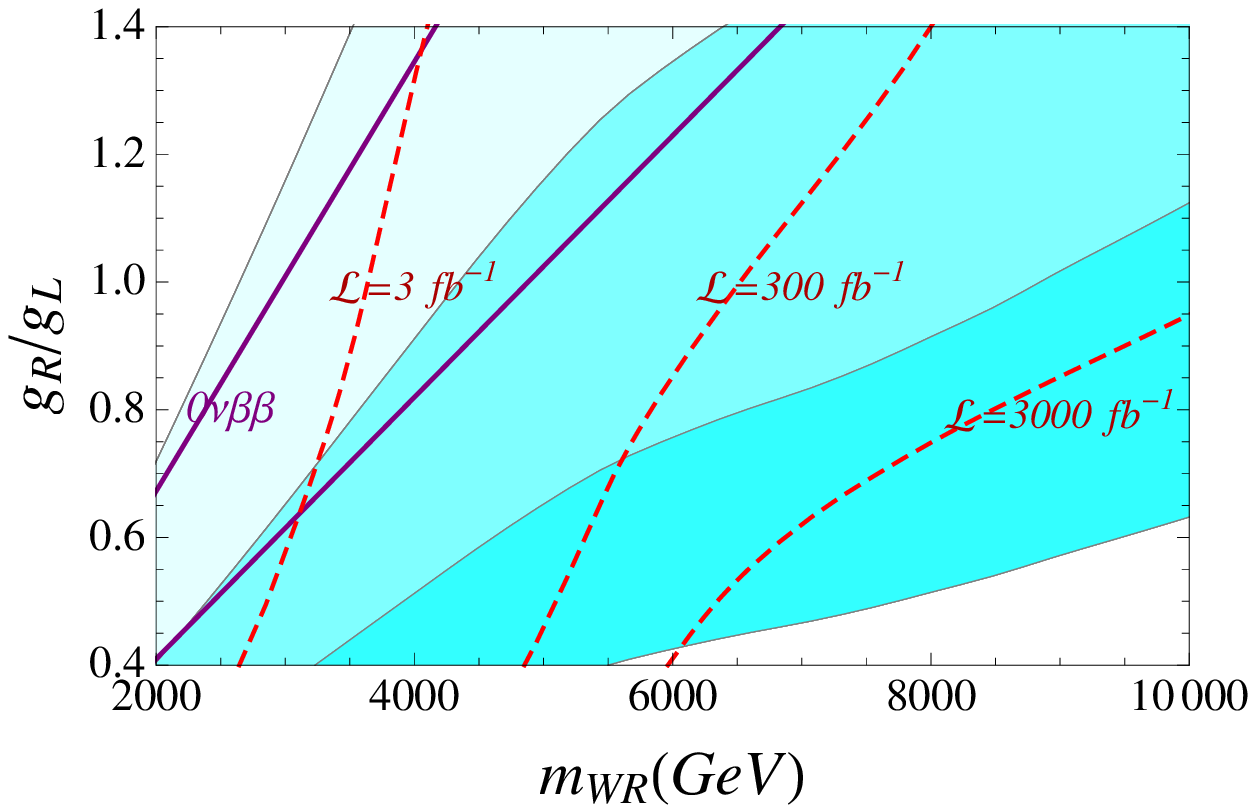}
\end{minipage}
\vspace{0.2 cm}
\caption{Constraints on LR model from double beta decay compared with
  parameter space where displaced vertex search at LHC will be
  sensitive. (a) top: $m_{W_R}$ vs $m_{N}$ for fixed $g_R = g_L$. (b)
  bottom: $g_R/g_L$ vs $m_{W_R}$ for fixed $m_{N} = 30GeV$.}
\label{fig:LR}  
\end{figure}

Unlike the sterile neutrino case, displaced vertex searches of a heavy
neutrino produced via a $W_R$ boson are not that sensitive on the cuts
used in the analysis. This can be understood from the fact that the
$W_R$ is much heavier than the $W_L$, and as a result, the $p_T$
distribution of the events extends to much higher $p_T$ then for the
sterile neutrino case.  Fig. (\ref{fig:LR}) shows two limits from
non-observation of $\znbb$. The weaker limit correspond to $T_{1/2} >
2.0 \times 10^{25}$ ys, while the line for larger values of $m_{W_R}$
is a expected sensitivity of $10^{27}$ ys. Again coloured bands show
the region that can be probed at the LHC in a 14 TeV run by a
displaced vertex search for decay lengths of order $L=1$ mm, $L=0.01$,
$0.1$ and $1$ m, with the largest lengths for
largest values of $m_{W_R}$. In this plot we have used the cuts $p_T >
40$ GeV and $|\eta| < 2.5$ for leptons and jets. Lines for different
assumed luminosities: ${\cal L} = 3$, $300$ and $3000$ fb$^{-1}$ are
shown. As demonstrated, LHC could probe so far unexplored ranges of
$m_N-m_{W_R}$ parameter space for luminosities as small as ${\cal L} =
3$ fb$^{-1}$ and for higher luminosities could be much more sensitive
than even (optimistic) future $\znbb$ results. Recall that current
direct search limits for $W_R$ at the LHC are of the order of
($2-2.5$) TeV for $m_{W_R}$, depending on luminosity used and search channel
\cite{ATLAS:2012ak,CMS:PAS-EXO-12-017}. The future 14 TeV run might be 
able to improve this limit to roughly ($4-4.5$) TeV \cite{Gninenko:2006br}. 
For $m_{W_R}$ much larger than this value, the $W_R$ will not be produced 
on-shell at the LHC.

Up to now, we have made the assumption that $g_R = g_L$ which is
assumed to be true in the minimal LRSM, but need not be so in general. In
Fig. (\ref{fig:LR}b) we thus plot, as in Fig. (\ref{fig:LR}a) a
comparison of $\znbb$ and LHC sensitivities but now in the plane
$g_R/g_L-m_{W_R}$ fixing $m_N = 30$ GeV. As is shown, especially for
small values of $g_R/g_L$ and large luminosity, the LHC can probe much
larger parts of the parameter space than even future $\znbb$ experiments.

\section{Summary}

We have discussed a possible displaced vertex search at the LHC 
to constrain (or discover!) signals of heavy neutral leptons. We 
have worked in the SM, extended by a sterile neutrino, and have 
also discussed the case of a left-right symmetric extension of the 
standard model.

Limits from non-observation of double beta decay are very 
stringent and LHC will need at least 300 fb$^{-1}$ to be 
competitive in the search for $|V_{e4}|^2$. However, the best 
limits on $|V_{\mu 4}|^2$, due to LEP, could be surpassed with 
order $(10-50$) fb$^{-1}$ already, although strongly depending on 
cuts as we have discussed in some detail. 

The prospects for discovery of a signal look much brighter in 
the left-right symmetric model. So far unexplored regions of 
parameter space may become asccesible with as little as $3$ fb$^{-1}$ 
and very large values of $m_{W_R}$ can be probed by our displaced 
vertex search for large luminosities.

Finally, we would like to close mentioning that we have assumed 
there is no significant generation mixing in the heavy neutrino sector, 
i.e. events are always $e^{\pm}e^{\pm}$- or $\mu^{\pm}\mu^{\pm}$-like. 
Obviously, and given the experience with the large mixing angles 
in the active neutrino sector, this does not have to be true at all. 
Particularly if there is one heavy neutrino which mixes significantly 
in the $\mu-\tau$ sector, limits from LHC will be much worse than 
our estimates for  $e^{\pm}e^{\pm}$- or $\mu^{\pm}\mu^{\pm}$-like 
events, due to the rather poor $\tau$ reconstruction efficiencies 
of the LHC experiments.

\medskip
\centerline{\bf Acknowledgements}


We thank V. Mitsou for various discusssions about the ATLAS detector.
This work was supported by UNILHC PITN-GA-2009-237920 and by the
Spanish MICINN grants FPA2011-22975, MULTIDARK CSD2009-00064, by the
Generalitat Valenciana (Prometeo/2009/091), by Fondecyt (Chile) under
grants 11121557, 1100582 and by CONICYT(Chile) project 791100017.
\appendix

\section{Heavy neutrino decay rates \label{A}}
Here we summarize the partial decay rates of the heavy sterile neutrino N in the mass range \mbox{1 GeV$\leq m_{N}\leq$ 80 GeV} analyzed in the present paper.  
In this range the semileptonic decays can be approximated by the decays into quark-antiquark pairs $N \rightarrow l(\nu) q_1 \bar q_2$. For more details see  
Ref. \cite{Helo:2010cw}. 

The partial decay rates are:
\begin{eqnarray}\label{lln-CC}
&&\Gamma(N\rightarrow l_1^{-}l_2^{+}\nu_{l_{2}} )=\\
\nonumber
&&= |V_{l_1 N}|^2
\frac{G_F^2}{192\pi^3} m_N^5 I_{1}(y_{l_1},y_{\nu_{l_{2}}}, y_{l_2})(1-\delta_{l_{1}l_{2}}), \\
\label{lln}
&&\Gamma(N\rightarrow \nu_{l_{1}}l_2^{-}l_2^{+} )=\\
\nonumber
&&=|V_{l_1 N}|^2 \frac{G_F^2}{96\pi^3} m_N^5
\left[\left(g^{l}_{L} g^{l}_{R}+ \delta_{l_{1}l_{2}}g^{l}_{R}\right) I_{2}(y_{\nu_{l_{1}}}, y_{l_{2}}, y_{l_{2}}) + \right. \\ \nonumber
&&\left.   + \left((g^{l}_{L})^{2} +(g^{l}_{R})^{2 }+ \delta_{l_{1}l_{2}} (1 +2 g^{l}_{L})\right) I_{1}(y_{\nu_{l_{1}}}, y_{l_{2}}, y_{l_{2}}) \right],\\
\label{3n}
&&\sum_{l_{2}=e,\mu,\tau}\Gamma(N\rightarrow \nu_{l_{1}} \nu_{l_{2}} \bar{\nu}_{l_{2}})= |V_{l_1 N}|^2 
\frac{G_F^2}{96\pi^3} m_N^5,\\
\label{lud}
&&\Gamma(N\to l_{1}^{-} u\bar{d})=\\
\nonumber
&&= |V_{l_{1}N}|^2\ |V_{u d}|^{2} \frac{G_F^2}{64\pi^3}m_N^5 
I_{1}(y_{l_{1}}, y_{u}, y_{d}),\\
 \label{nuqq}
&&\Gamma(N\to \nu_{l_{1}}\, q\bar{q}) = \\
\nonumber
&&=|V_{l_{1}N}|^2\frac{G_F^2}{32\pi^3}m_N^5 \left[ g^{q}_{L} g^{q}_{R} I_{2}(y_{\nu_{l_{1}}}, y_{q}, y_{q}) + \right. \\ \nn
&&\left. + \left((g^{q}_{L})^{2} +(g^{q}_{R})^{2 })\right) I_{1}(y_{\nu_{l_{1}}}, y_{q}, y_{q}) \right].
\end{eqnarray} 
Here we denoted $y_{i} = m_{i}/m_{N}$ with $m_{i} = m_{l}, m_{q}$ being the lepton and quark masses, respectively. For the quark masses we use the values 
\mbox{$m_{u}\approx m_{d} = 3.5$ MeV},  \mbox{$m_{s} = 105$ MeV}, \mbox{$m_{c} = 1.27$ GeV}, \mbox{$m_{b} = 4.2$ GeV.}  In Eqs. (\ref{lud}), (\ref{nuqq})  
we denoted  $u=u,c,t$; $d=d,s,b$ and $q=u,d,c,s,b,t$. The SM neutral current couplings of leptons and quarks are\\[-25mm]
\begin{eqnarray}\label{NC-coupl}
g^{l}_{L} &=&-1/2 + \sin^2\theta_W, \ \  \ \ \ \ \ \ \ g^{l}_{R} = \sin^2\theta_W,\\  
\nonumber
g^{u}_{L}&=& 1/2 - (2/3) \sin^2\theta_W,  \ \  \ \ g^{u}_{R}= -(2/3)\sin^2\theta_W,\\
\nonumber
g^{d}_{L} &=& -1/2 + (1/3) \sin^2\theta_W, \ \  g^{d}_{R}= (1/3)\sin^2\theta_W.
\end{eqnarray}\\[-25mm]
The kinematical functions are:\\[-25mm]
\begin{eqnarray}\label{kin-fun-1}
 &&I_{1}(x,y,z)= \\
 \nonumber
 &&=12 \int\limits_{(x+y)^{2}}^{(1-z)^{2}} \frac{ds}{s}
(s-x^2-y^{2})(1+z^2-s) \times \\
\nonumber
&&\hspace{30mm}\times \lambda^{1/2}(s, x^{2}, y^2) \lambda^{1/2}(1, s, z^2),\nn
\\ 
&& I_{2}(x,y,z)= \\
\nonumber
&&=24 y z \int\limits_{(y+z)^{2}}^{(1-x)^{2}} \frac{d s}{s} (1+x^{2}-s)
\lambda^{1/2}(s,y^{2},z^{2})\lambda^{1/2}(1,s,x^{2}).
\end{eqnarray}\\[-25mm]
The total decay rate $\Gamma_{N}$  of the heavy neutrino $N$ is:\\[-25mm]
\begin{eqnarray}\label{total-4}
\nonumber
\Gamma_{N}& =&  \sum_{l_{1}, l_{2}, {\cal H}}\left[ 2  \Gamma(N\rightarrow l_{1}^{-}{\cal H}^{+}) +
2  \Gamma(N\rightarrow l_{1}^{-} l_{2}^{+}\nu_{l_{2}}) +\right. \\ 
&+&\left.\Gamma(N\rightarrow \nu_{l_{1}}{\cal H}^{0}) + \Gamma(N\rightarrow l_{2}^{-} l_{2}^{+}\nu_{l_{1}})\right. + \\
\nonumber
&+&\left.\Gamma(N\rightarrow \nu_{l_{1}}\nu_{l_{2}} \bar{\nu}_{l_{2}}) \right],
\end{eqnarray}\\[-25mm]
where we denoted ${\cal H}^{+} = \bar{d} u,  \bar{s} u,  \bar{d} c,  \bar{s} c$ and  ${\cal H}^{0} = \bar{q} q$. 
The factor 2 of the first two terms is due to Majorana nature of heavy sterile neutrino $N$ studied in the present paper.
This factor is related with the fact that both charge conjugate final states are allowed: 
$N\rightarrow l_{1}^{-}l_{2}^{+} \nu_{l_{2}}, l_{1}^{+}l_{2}^{-} \bar{\nu}_{l_{2}}$ and $N\rightarrow l^{\mp}{\cal H}^{\pm}$.    
\bibliographystyle{h-physrev5}

\end{document}